%% file: root.tex
\documentclass[a4paper, 10pt, conference]{ieeeconf}      

\IEEEoverridecommandlockouts                              

\usepackage{cite}
\usepackage{amsmath,amssymb,amsfonts}
\usepackage{algorithmic}
\usepackage{graphicx}
\usepackage{textcomp}
\usepackage{gensymb}
\usepackage{xcolor}
\usepackage{color}		
\usepackage{multirow}		
\usepackage{array}										
\usepackage{colortbl}
\usepackage{siunitx}
\usepackage[english]{babel}	
\usepackage{multirow}
\usepackage[hidelinks]{hyperref}    
\usepackage{blindtext}
\usepackage[ruled, linesnumbered]{algorithm2e}

\title{\LARGE \bf
Learning-based Position and Stiffness Feedforward Control of \\Antagonistic Soft Pneumatic Actuators using Gaussian Processes
}

\author{Tim-Lukas Habich, Sarah Kleinjohann and Moritz Schappler$^{1}$%
\thanks{$^{1}$All authors are with the Leibniz University Hannover, Institute of Mechatronic Systems, 30823 Garbsen, Germany,
	{\tt {tim-lukas.habich@imes.uni-hannover.de}}}%
}

\newif\ifcopyright
	\copyrighttrue

\newcommand{\mm}[1]{\boldsymbol{#1}}
 
\newcommand{\sk}[5]{{}_{\mathrm{#2}}^{\mathrm{#3}}{#1}^{\mathrm{#4}}_{\mathrm{#5}}}

\newcommand{\e}[2]{\begin{equation} #1 \label {eq:#2} \end{equation}}
\newcommand{\mc}[1]{\mathcal{#1}}
\newcommand{\ind}[1]{\mathrm{#1}}
\usepackage[mathscr]{euscript}

\definecolor{Gray}{gray}{0.85}

\newcolumntype{M}[1]{>{\centering\arraybackslash}m{#1}}
\newcolumntype{N}{@{}m{0pt}@{}}
\makeatletter
\newcommand{\removelatexerror}{\let\@latex@error\@gobble}
\makeatother

\begin{document}
\ifcopyright
{\LARGE IEEE Copyright Notice}
\newline
\fboxrule=0.4pt \fboxsep=3pt

\fbox{\begin{minipage}{1.1\linewidth}  
		\textcopyright\,\,2023\,\,IEEE. Personal use of this material is permitted. Permission from IEEE must be obtained for all other uses, in any current or future media, including reprinting/republishing this material for advertising or promotional purposes, creating new collective works, for resale or redistribution to servers or lists, or reuse of any copyrighted component of this work in other works. \\
		
		Accepted to be published in: Proceedings of the IEEE-RAS International Conference on Soft Robotics (RoboSoft), April~3 -- April~7, 2023, Singapore.\\
		
		DOI: 10.1109/RoboSoft55895.2023.10122057
		
\end{minipage}}
\else
\fi
\graphicspath{{./images/}}

\maketitle
\thispagestyle{empty}
\pagestyle{empty}

\begin{abstract}
Variable stiffness actuator (VSA) designs are manifold. Conventional model-based control of these nonlinear systems is associated with high effort and design-dependent assumptions. In contrast, machine learning offers a promising alternative as models are trained on real measured data and nonlinearities are inherently taken into account. Our work presents a universal, learning-based approach for position and stiffness control of soft actuators. After introducing a soft pneumatic VSA, the model is learned with input-output data. For this purpose, a test bench was set up which enables automated measurement of the variable joint stiffness. During control, Gaussian processes are used to predict pressures for achieving desired position and stiffness. The feedforward error is on average $11.5\%$ of the total pressure range and is compensated by feedback control. Experiments with the soft actuator show that the learning-based approach allows continuous adjustment of position and stiffness without model knowledge.
\end{abstract}
\section{Introduction}
Intraluminal procedures within minimally invasive surgery (MIS) offer a reduction of incisions resulting in less postoperative pain and faster recovery of the patient. Thus, MIS provides a sustainable improvement of medical procedures by access via natural orifices. Two main requirements for endoscopes exist: high \textit{maneuverability} to locomote into the target area and sufficient \textit{stiffness} within the target area for tissue manipulation. Conventional endoscopes cannot achieve these goals. Rigid ones provide high stiffness but no maneuverability and flexible endoscopes have low stiffness at the actuated distal end due to the shaft's passive structure. By replacing this passive shaft with a fully actuated kinematic chain, continuum robots \cite{BurgnerKahrs.2015} or hyper-redundant (discrete) snake robots \cite{Chirikjian.1992} can be built to meet both demands. Besides endoscopy with very strict space requirements, small snake robots are promising for several medical procedures -- these could e.g. act within computed tomography (CT) scanners realizing robotic surgery under image guidance.

In all of the named applications, there is direct contact between robot and humans. Accordingly, there are high safety requirements making soft robots ideally suited. However, the \textit{compliant system} used to manipulate tissue should provide the operator with a \textit{stiff working platform}. What seems to be a contradiction at first glance is ubiquitous in nature: Vertebrates like humans \cite{Hogan.1984} but also invertebrates such as the octopus \cite{Althoefer.2018} can change the stiffness of the limbs by antagonistic arrangement of the muscles. Consequently, the design of a soft snake robot with variable stiffness enables both safe human-robot interaction and a stable working platform for manipulation when needed. Due to the significant power density of pneumatic actuators, they have a high miniaturization potential compared to miniaturized electric motors. In addition, the use of soft pneumatic robots in CT or magnetic resonance imaging (MRI) scanners is uncritical. Our long-term goal is therefore to build a modular snake robot consisting of pneumatic actuators with variable stiffness. Model-based simultaneous position and stiffness control of such rubber-like robots is challenging due to complex geometries, strong material nonlinearities and air compressibility \cite{Xavier.2022}. Nonlinear deformations of viscoelastic (time, temperature and frequency dependent) material are difficult to describe with conventional models. This work focuses on an alternative approach: position and stiffness control using machine learning. 

	\begin{figure}[tbp]
		\centering
		\resizebox{1\linewidth}{!}{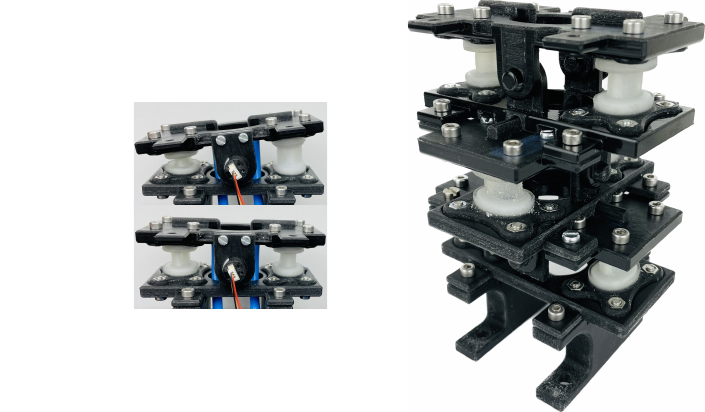}
		\caption{Soft pneumatic actuator for snake-robot applications: (a)~Both continuous change of joint angle (upper two images) and increase of joint stiffness at constant position by antagonistic coactivation (lower two images) are possible. (b)~The actuator is modular in terms of stackability. Three units exemplarily form an articulated soft robot with variable compliance.} \label{fig:cover}
\end{figure}
\subsection{Related Work}
Soft robots can be classified as soft continuum robots and articulated soft robots (ASRs) \cite{GeorgeThuruthel.2018}. The latter have compliant joints within a vertebrate-like structure consisting of rigid links as the one shown in Fig.~\ref{fig:cover}(b) and are the focus of our work. The advantages of ASRs are the simple mounting of rotational encoders and kinematics modeling using classical methods for rigid robots. The introduction of controlled compliance is possible using VSAs~\cite{Wolf.2016} which are a subset of variable impedance actuators \cite{Vanderborght.2013}.

Several model-based methods to control robots with elastic joints exist: feedback linearization \cite{Luca.1998}, LQR-based gain scheduling \cite{Sardellitti.2012}, finite element method \cite{Koehler.2019}, adaptive control \cite{Tonietti.2002,Trumic.2021} using Chou-Hannaford model of pneumatic artificial muscles~\cite{Chou.1996} and simultaneous position and stiffness control using sliding mode control or model predictive control \cite{Best.2021}. As mentioned above, deriving analytical models required in all publications is challenging for soft pneumatic actuators and only possible with several design-dependent assumptions. Various designs of pneumatic VSAs can be found, e.g. an antagonistic pneumatic artificial muscle with internal and external chambers \cite{Usevitch.2018}, an antagonistic soft pneumatic actuator with casted bellows and wrapped fibers~\cite{Skorina.2015} or a variable stiffness soft module with four degrees of freedom (DoF) consisting of tendons, two molded air chambers and wrapped fibers \cite{Manfredi.2018}. This variety of VSA designs, which are difficult to model analytically, leads us to the main question of this paper: Is it possible to control position and stiffness using a \textit{design-independent approach without model knowledge} which is applicable to several VSAs?

Controlling robots without the need to model the system is possible using \textit{data-driven} approaches. Within soft robotics, this field is still in its infancy \cite{Wang.2021b}. There are only a few publications dealing with learning-based position and stiffness control of VSAs. Ansari et al. \cite{Ansari.2018} use cooperative multi-agent reinforcement learning in order to optimize stiffness and position simultaneously. This requires a high number of samples with optimal and suboptimal state-action pairs which is difficult to generate on the test bench. Data-driven position and stiffness control was successfully realized for a single actuator by using long short-term memory units~\cite{Luong.2021}. However, special (twisted-coiled polymer) actuators were considered. Their stiffness is closely related to the temperature. Thus, position and temperature were controlled which are both \textit{measurable} quantities during feedback control. For arbitrary actuators, a model is required that describes the current stiffness as a function of the control variables (e.g. pressures or tendon forces). 

Within \cite{Angelini.2018,Mengacci.2021}, iterative learning control (ILC) is applied to realize model-free position and stiffness control of ASRs. However, the physical characteristics of the elastic robot joints are required and are taken from a data sheet \cite{Grioli.2015} which consists of theoretical mathematical models calibrated with field data. In \cite{Lukic.2016}, a neural network is used to implement feedforward control of motor positions given desired position and stiffness references. However, the stiffness model is also taken from a data sheet of the VSA. Hofer and D'Andrea~\cite{Hofer.2020} use ILC to compensate for repetitive disturbances such as the interaction between two angles of a universal joint and other unmodeled effects like hysteresis behavior of the material and joint friction. However, only three discrete joint stiffnesses are used (soft, medium, stiff). They conclude that it should be investigated how the joint stiffness can be continuously controlled. This is the core of our work.

\subsection{Contributions}
Della Santina et al. \cite{DellaSantina.2017} show that high-gain feedback control causes soft actuators to lose their compliance during external contact. Relatively small feedback gains are required to preserve the desired compliance. In order to still achieve good tracking performance, an accurate model is required for feedforward control of necessary actuator inputs. An approach widely used in rigid robot control has so far found limited application in stiffness control of soft robots: \textit{model learning} \cite{NguyenTuong.2011}. Due to training with measured data, the learned models inherently include unknown nonlinearities that are difficult to describe using classical modeling approaches. Since the approach should be universal for arbitrary VSAs, non-parametric regression approaches are ideally suited. We use Gaussian processes (GPs)~\cite{Rasmussen.2008} which have the additional advantage of determining prediction uncertainties. 

Our contributions are therefore: 
\begin{enumerate}
	\item Presenting a 3D-printed, modular soft pneumatic actuator with variable stiffness for snake-robot applications, 
	\item building a test bench and automated recording of a vast amount of VSA input-output data in order to
	\item train Gaussian processes to predict necessary pressures given desired joint angle and stiffness to
	\item realize simultaneous, learning-based position and stiffness control of soft actuators.
\end{enumerate}
This is the first time that Gaussian processes have been used for stiffness control of soft actuators. Further, the learning-based approach is universally applicable to arbitrary VSAs without model knowledge. The paper is structured as follows: Section~\ref{main_actuator} describes our soft pneumatic actuator. Based on this, learning-based control using Gaussian process regression is presented (Sec.~\ref{main_control}), followed by experiments (Sec.~\ref{experiments}) and conclusions (Sec.~\ref{conclusions}).
\section{Modular Soft Pneumatic Actuator with Variable Stiffness}\label{main_actuator}
The soft pneumatic actuator used in this work is modular and can be stacked together to form a spatial snake robot. For this purpose, the actuators are mounted by successive rotation of $\SI{90}{\degree}$ which is exemplarily illustrated in Fig.~\ref{fig:cover}(b) for a robot consisting of three actuators. This section describes the actuation principle (\ref{princ_act}), the manufacturing process and the assembly of the single actuator (\ref{assem_act}).
\subsection{Actuation Principle}\label{princ_act}
The VSA is built from two soft air bellows acting antagonistically on a single joint. Thus, the variable joint angle $q(p_1,p_2)$ depends on the bellows pressures of agonist $p_1$ and antagonist $p_2$. This redundancy causes various pressure combinations to lead to the same joint angle and allows to additionally adjust the joint stiffness $s(p_1,p_2)$. In general, stiffness can be varied by changing the spring preload or the transmission ratio between the output and the spring as well as by influencing the physical spring properties \cite{Wolf.2016}. For our design, the stiffness variation is reached by changing the preload of the nonlinear springs (soft bellows). 

The main principle is presented in Fig.~\ref{fig:cover}(a). Whereas net joint stiffness is determined by the sum of the activities of agonist and antagonist muscle groups (antagonist coactivation), the net torque results predominantly from the difference of both activities \cite{Hogan.1984}. Therefore, changing the difference of bellows pressures 
\e{\Delta p=p_1-p_2}{pressure_dif}
enables the variation of the joint angle. In contrast, a higher pressure level 
\e{\bar{p}=\frac{p_1+p_2}{2}}{pressure_mean}
at constant pressure difference realizes a stiffening at constant joint angle. Stiffening is also theoretically possible when the joint angle limits are reached. At these operating points, the joint stiffness is highly asymmetric. Stiffness control near the joint limits is outside the scope of this paper.

\subsection{Manufacturing and Assembly}\label{assem_act}
\begin{figure}[bp]
		\centering
		\resizebox{1\linewidth}{!}{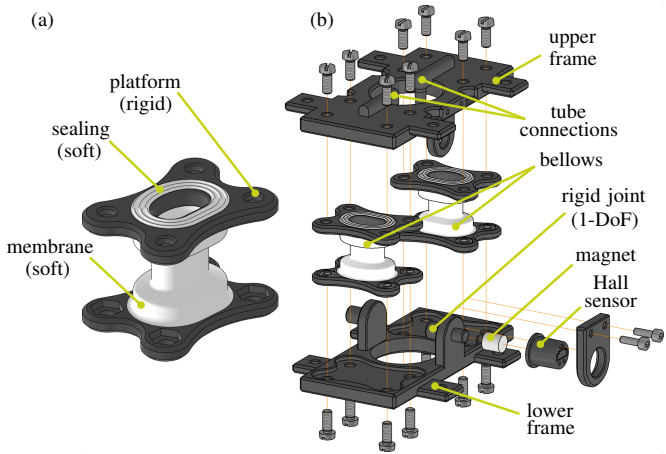}
		\caption{Soft actuator components: (a) The printed bellows consists of a soft membrane, a rigid platform for mounting and a sealing made of soft material. (b) Each actuator consists of two antagonistically acting bellows which are mounted on the upper and lower half of the frame with screws. The joint angle is measured with a Hall sensor. The printed tube connections are located inside the actuator.} \label{fig:construction_view}
\end{figure}

One modular actuator unit, shown in Fig.~\ref{fig:construction_view}(b), comprises a rigid frame, a Hall sensor for absolute angle detection and two soft air bellows. 
The frame is split into an upper and a lower part, each of them with a hole in the center to create a passage for the air supply tubes inside the robot. Printed connections are used to attach the tubes to the upper frame by means of a press fit. The bellows are supplied with compressed air via a printed channel within the frame.

Each bellows consists of two rigid platforms with the soft membrane in between them which is illustrated in Fig.~\ref{fig:construction_view}(a). These platforms allow the bellows to be mounted with screws onto the rigid frame and enable the replacement of bellows in case of material failure during use. This can happen due to fatigue of the soft material caused by the permanent stretching and compressing of the membrane. To reach optimal compromise between flexibility and persistence, the membrane has a thickness of $\SI{1.75}{\milli\meter}$ and a height of $\SI{23}{\milli\meter}$. At $q{=}\SI{0}{\degree}$, the distance between upper and lower half of the frame is $\SI{3}{\milli\meter}$ less than the bellows height. Consequently, the bellows are mounted in compressed states. This is advantageous as material stresses in the soft membrane are reduced under elongation during movement and material failure occurs significantly later. A leak-tight connection after fastening the bellows to the upper frame is ensured by a soft sealing lip which surrounds the hole for air supply.

To realize a snake robot with hollow shaft, the joint shaft is split. The lower frame includes the split shaft, while the upper part contains the bearings. Both parts are assembled by sliding the bearings over the shaft. In order to measure the joint angle, a magnet is placed on a small shaft extension on the side of the actuator. Afterwards, the corresponding hall sensor is screwed to the upper frame and simultaneously secures the frame from sliding apart.

Frame and bellows are 3D-printed with Polyjet technology\footnote{Stratasys Objet350 Connex3}, whereby printing soft and rigid material during one printing process is possible. As materials, the rigid VeroBlackPlus and the flexible Agilus30 are used. Hollow and protruding structures such as the tube connections or the bellows can be printed precisely since cavities are filled with support material that can be removed after printing. The removal happens manually with dentist tools, but smaller residuals of support material can be loosened in an ultrasound bath and rinsed with water afterwards.
\section{Simultaneous Position and Stiffness Control using Gaussian Processes}\label{main_control}
After presenting the pneumatic VSA, the learning-based control is introduced in the following (\ref{learn_based_con}). Basics of Gaussian process regression are first described (\ref{pre_GP}).
\subsection{Preliminaries on Gaussian Process Regression}\label{pre_GP}
A Gaussian process 
\e{f(\mm{x})\sim\mc{GP}(m(\mm{x}),k(\mm{x},\mm{x}'))}{gp} is a set of random variables $f_i{=}f(\mm{x}_i)$ for which each finite subset follows a Gaussian distribution \cite{Rasmussen.2008}. It is defined by the mean $m(\mm{x})$ and covariance function $k(\mm{x},\mm{x}')$. These can be chosen to incorporate prior knowledge of the system. We assume zero-mean ($m(\mm{x}){\equiv}0$) since the outputs can be normalized to zero-mean. In terms of covariance, the widely used squared exponential (SE) kernel 
\e{k(\mm{x},\mm{x}')=\sigma_{f}^2\exp{(-\frac{1}{2l^2}||\mm{x}-\mm{x}'||^2_2)}}{SE}
with length-scale $l$ and signal variance $\sigma_{f}^2$. Further, \mbox{$\mm{X}{=}(\mm{x}_1,...,\mm{x}_n)^\ind{T}$} describes the inputs and \mbox{$\mm{y}{=}(y_1,...,y_n)^\ind{T}$} the related noisy observations. The noise in the system \mbox{$\epsilon_i{\sim}\mc{N}(0,\sk{\sigma}{}{}{2}{n})$} is assumed to be Gaussian with zero-mean and noise variance $\sk{\sigma}{}{}{2}{n}$. Therefore, the relation 
\e{y_i=f_i+\epsilon_i}{datapoint}
applies to a data point $i$. The random variable $f_i$ describes the model to be learned which -- with the additive noise component -- is equal to the noisy observation $y_i$. 

For a known training set $\mm{\mc{D}}{=}\{\mm{X},\mm{y}\}$ with $n$ data points, there is a regression problem with the aim to predict $f_*$ at a query point $\mm{x}_*$. Incorporating this knowledge $\mm{\mc{D}}$ about the searched model 
is done by conditioning the joint Gaussian prior distribution on the observations. This results in the predictive distribution of Gaussian process regression \cite{Rasmussen.2008}
\e{f_*|\mm{X},\mm{y},\mm{x}_*\sim\mc{N}(\mu_*,\sigma_*)}{predictive_distribution}
consisting of predicted mean
\e{\mu_*=\mm{k}_*^\mathrm{T}\mm{\alpha}\quad\mathrm{with}\quad\mm{\alpha}=(\mm{K}+\sigma_\mathrm{n}^2\mm{I})^{-1}\mm{y}}{mean}
and corresponding uncertainty
\e{\sigma_*=k_*-\mm{k}_*^\mathrm{T}(\mm{K}+\sigma_\mathrm{n}^2\mm{I})^{-1}\mm{k}_*}{siga_star}
using compact kernel notation: \mbox{$\mm{K}{=}\mm{K}(\mm{X},\mm{X}){\in}\mathbb{R}^{n\times n}$}, \mbox{$\mm{k}_*{=}\mm{k}(\mm{X},\mm{x}_*){\in}\mathbb{R}^{n\times 1}$}, \mbox{$k_*{=}k(\mm{x}_*,\mm{x}_*){\in}\mathbb{R}$}. Learning is carried out by adjusting the hyperparameters $\{\sk{\sigma}{}{}{2}{n},\sigma_{f}^2,l\}$ to the dataset. This is done by maximizing the marginal likelihood $p(\mm{y})$. The library GPy \cite{GPy.2012} is used for training. 
\subsection{Learning-based Control}\label{learn_based_con}
Within the position and stiffness control of the antagonistic actuator from Sec.~\ref{main_actuator}, the controlled variables are the joint angle $q$ and the joint stiffness $s$. At steady-state, these are independently adjustable by means of the bellows pressures $\mm{p}{=}(p_1,p_2)^\mathrm{T}$. In order to design a suitable feedforward control of desired pressures $\mm{p}_\mathrm{d}{=}(p_\mathrm{1,d},p_\mathrm{2,d})^\mathrm{T}$, the relation
\e{\mm{p}=\mm{f}(q,s)=\left( \begin{array}{c}
		f_\mathrm{I}(q,s)\\
		f_\mathrm{II}(q,s)\end{array}\right)}{desired_model}
must be known. Each bellows pressure requires a model $f_\mathrm{I}(q,s)$ and $f_\mathrm{II}(q,s)$, respectively. This relation is learned by means of GP regression with inputs $\mm{x}{=}(q,s)^\mathrm{T}$ and output $y_\mathrm{I}{=}p_1$ or $y_\mathrm{II}{=}p_2$. Accordingly, a total of two Gaussian processes are trained. The feedforward pressures 
\e{\mm{p}_\ind{FF}=\left( \begin{array}{c}
		p_\ind{1,FF}\\
		p_\ind{2,FF}\end{array}\right)=\mm{\mu}_*=\left( \begin{array}{c}
		\mm{k}_*^\mathrm{T}\mm{\alpha}_\ind{I}\\
		\mm{k}_*^\mathrm{T}\mm{\alpha}_\ind{II}\end{array}\right)}{p_ff}
are obtained by (\ref{eq:mean}) at a query point $\mm{x}_*{=}(q_\ind{d},s_\ind{d})^\ind{T}$ given desired position $q_\ind{d}$ and joint stiffness $s_\ind{d}$. During control, the Woodbury vectors $\mm{\alpha}_\ind{I},\mm{\alpha}_\ind{II}$ remain constant. In each cycle, only the covariance vector $\mm{k}_*$ has to be recalculated.

To compensate for model errors, an additional feedback loop is closed. The main problem here is that \textit{joint stiffness cannot be measured during operation}. This is due to the fact that joint stiffness, or impedance in general, is a differential operator relating the time course of measurable quantities (torque and angle) \cite{Grioli.2010}. A feedback of the variable stiffness is therefore only possible with observer approaches, which are out of the paper's scope. Our work is only intended to answer the question whether it is possible to build up a stiffness control with a black-box approach. We therefore assume that the learned model for feedforward control is sufficiently accurate and only feed back the joint angle. This assumption is discussed further in Sec.~\ref{conclusions}.

A proportional-integral (PI) controller with low gains $K_\ind{P},\,K_\ind{I}$ is used within the feedback loop in order to minimally influence the desired compliance in case of contact. As already mentioned, the difference of agonistic and antagonistic muscle group $\Delta p$ influences the joint angle $q$. Accordingly, the controller output $\Delta p_\ind{FB}{=}p_\ind{1,FB}{-}p_\ind{2,FB}$ is calculated using position error.

\begin{figure}[tbp]
	\vspace{2mm}
	\centering
	\resizebox{1\linewidth}{!}{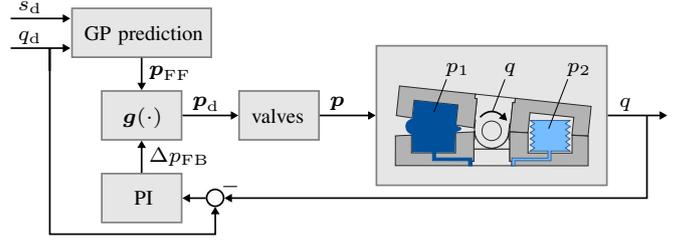}
	\caption{Block diagram of the proposed learning-based position and stiffness control of the soft actuator: The feedforward pressures $\mm{p}_\ind{FF}$ result from the Gaussian processes (\ref{eq:desired_model}) and are added to the target pressure difference from the feedback loop $\Delta p_\ind{FB}$ by means of (\ref{eq:p_d}).} \label{fig:blockschaltbild}
	\vspace{-2mm}
\end{figure}

The block diagram of the proposed control system is illustrated in Fig.~\ref{fig:blockschaltbild}. Merging of the feedforward pressures with the feedback term is done by means of
\e{\mm{p}_\ind{d}=\mm{g}(\mm{p}_\ind{FF},\Delta p_\ind{FB})=\mm{p}_\ind{FF}+\frac{\Delta p_\ind{FB}}{2}(1,-1)^\ind{T}.}{p_d}
Accordingly, the pressure difference from the feedback loop for fine position adjustment is equally divided between the two bellows. The desired pressures $\mm{p}_\ind{d}$ are passed to the valve-internal pressure controllers as targets and current pressures $\mm{p}$ finally act on the controlled system (VSA).

One final remark has to be made: Due to the use of the Euclidean distance within the SE kernel, inputs are scaled to zero mean and unit variance before GP training. The same is performed for the outputs due to our assumption ($m(\mm{x}){\equiv}0$) in \ref{pre_GP}. Note that during operation, the inputs $(q_\ind{d},s_\ind{d})$ must be scaled as well and the predicted outputs must be rescaled to obtain feedforward pressures $\mm{p}_\ind{FF}$. This standardization is performed within the ``GP prediction" block in Fig.~\ref{fig:blockschaltbild}.
\section{Experiments}\label{experiments}
The proposed learning-based position and stiffness control is validated using the 3D-printed VSA. For this purpose, the test bench is first introduced (\ref{test_bench}) and the automated recording of the training data is presented (\ref{recording_data}). Prediction accuracy is examined using cross-validation (\ref{cv_sec}). Finally, an experiment with the actuator shows that the suggested approach allows simultaneous adjustment of position and stiffness \textit{without model knowledge} (\ref{control_sec}).
\subsection{Test Bench}\label{test_bench}
Figure~\ref{fig:test_bench} illustrates the test bench built to record input-output data. The setup consists of the following components.
\subsubsection{Pneumatics}
After the supply unit consisting of pressure supply, shut-off valve and filter regulator, a proportional valve is installed for each bellows. A 3-way piezo valve\footnote{Festo VEAA-B-3-D2-F-V1-1R1, resolution: $\SI{5}{mbar}$} with integrated pressure control is used. The valve's sensor measures the pressure $p_\diamond$ at the working port ($2$) and the valve's controller adjusts it to the set point $p_{\diamond\mathrm{,d}}$.
\subsubsection{VSA}
The actuator is equipped with a Hall-effect absolute encoder\footnote{Megatron MAB12AH, resolution: $\SI{0.35}{\degree}$}. To measure the variable stiffness, a torque sensor\footnote{Weiss Robotics KMS40, resolution: $\SI{0.001}{Nm}$} with motor\footnote{Beckhoff AM8131, resolver resolution: $\SI{0.03}{\degree}$} is connected via rigid connections.
\subsubsection{Communication}
Test-bench communication is done via EtherCAT protocol and the corresponding open-source tool EtherLab\footnote{\tt\url{https://www.etherlab.org/}} which was modified with an external-mode patch and a shared-memory real-time interface\footnote{\tt \url{https://github.com/SchapplM/etherlab-examples}}. Accordingly, the robot operating system (ROS) package of the torque sensor\footnote{\tt \url{https://github.com/ipa320/weiss\_kms40}} can be integrated into the real-time communication to read the torque value $\tau$ at $\SI{500}{Hz}$ (not hard real-time capable). The control design is performed on the development computer (DEV-PC) using Matlab/Simulink. The compiled model is then run on the real-time computer (RT-PC). Sensor values (pressures $\mm{p}$, joint angle $q$, motor position $q_\ind{m}$) and commands (desired pressures $\mm{p}_\ind{d}$, desired motor position $q_\ind{m,d}$) are read in or set via the EtherCAT real-time bus with input and output terminals\footnote{Beckhoff EL3702, EL4102, EL7211-0010}, respectively. The setup allows real-time data acquisition and control with a cycle time of $\SI{1}{\milli\second}$ except the reading of the torque values.

	\begin{figure}[tbp]
		\centering
		\resizebox{1\linewidth}{!}{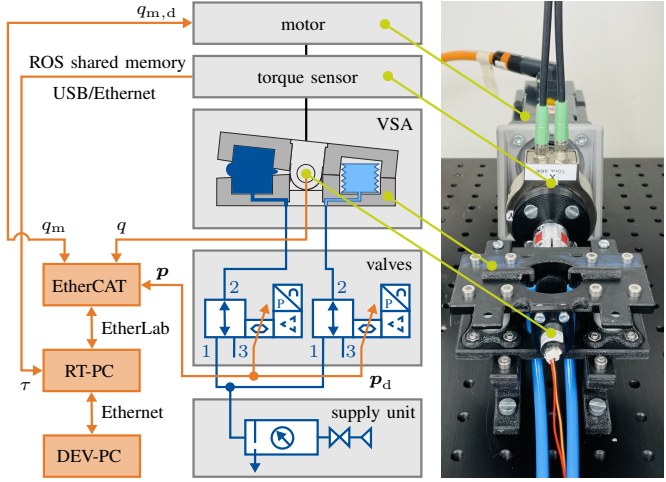}
		\caption{Test-bench architecture for measuring variable joint stiffness: Orange color represents communication elements such as sensor cables and blue color represents pneumatic components. The soft actuator with encoder is connected to the torque sensor and the motor via rigid connections.} \label{fig:test_bench}
\end{figure}
\subsection{Automated Recording of Training Data}\label{recording_data}
Joint stiffness can be varied by bellows pressures and is measured for an arbitrary number of pressure combinations to generate a dataset. For each data point $i$ consisting of inputs $\mm{x}_i{=}(q_i,s_i)^\ind{T}$ and outputs $y_{\ind{I},i}{=}p_{1,i}$ or $y_{\ind{II},i}{=}p_{2,i}$, the following steps must be performed:
\begin{enumerate}
	\item pass desired pressures $\mm{p}_{\ind{d,}i}$ to pressure controllers
	\item wait $\SI{5}{\second}$ allowing actuator to reach steady-state angle $q_i(p_{1,i},p_{2,i})$ \textit{without} external forces from the motor
	\item set new origin $\tilde{q}_\ind{m}{=}q_\ind{m}{-}q_i{=}\SI{0}{\degree}$ at steady-state
	\item drive motor trajectory between $\SI{-1}{\degree}{<}\tilde{q}_\mathrm{m}{<}\SI{1}{\degree}$ and acquire torques $\tau$ with torque sensor
\end{enumerate}
Note that within step {2)}, the motor still rotates by the same angle as the soft actuator due to the rigid connection between motor and soft actuator. In contrast, the motor forces the actuator to follow the same trajectory during step {4)}.
\begin{figure}[tbp]
	\vspace{1.5mm}
	\centerline{\includegraphics[width=\linewidth]{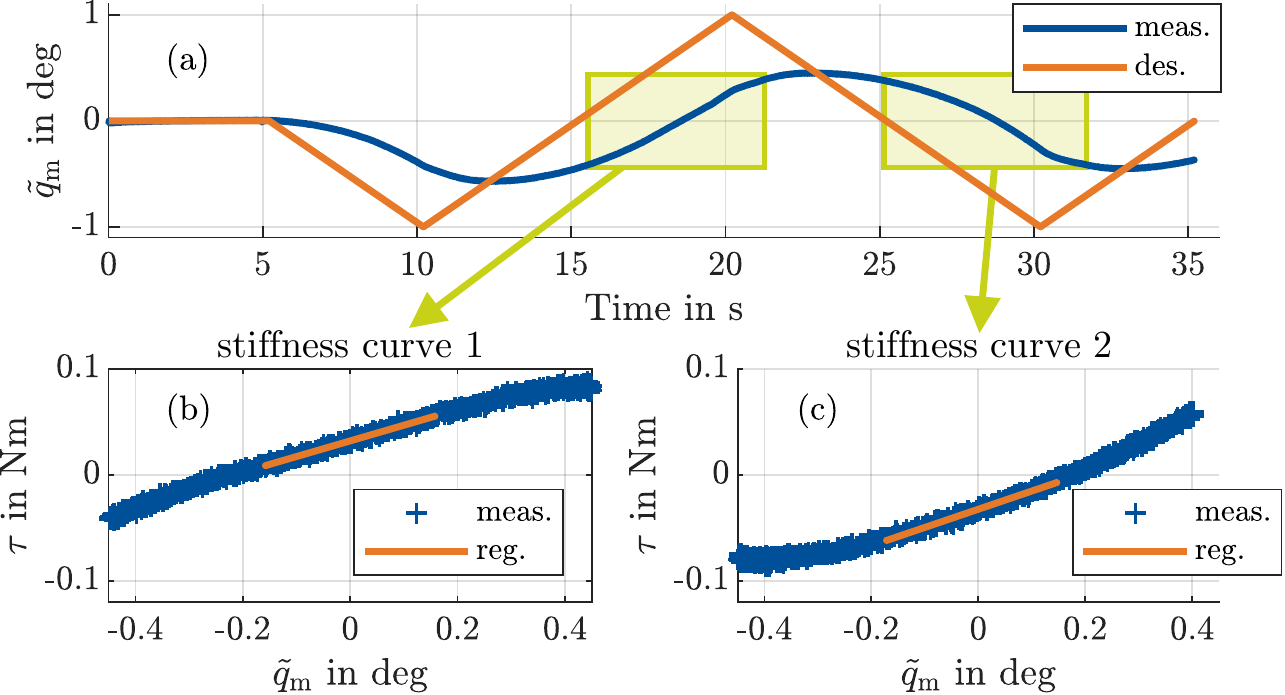}}
	\caption{Automated measurement of joint stiffness at arbitrary steady-state angle $q_i$: (a) A trajectory is driven between $\SI{-1}{\degree}{<}\tilde{q}_\mathrm{m}{=}q_\ind{m}{-}q_i{<}\SI{1}{\degree}$. (b)--(c)~During the two zero crossings in (a), the torques are recorded using torque sensor and two stiffness curves result. To obtain the stiffness measure $s_i$ for data point~$i$, the average of the two slopes is calculated.}
	\label{fig:trajectory}
	\vspace{-1.5mm}
\end{figure}

The desired and measured trajectory of the motor is shown in Fig.~\ref{fig:trajectory}(a) during an exemplary acquisition of a data point. The default position control of the motor is used. A considerable deviation between the target and the measured motor trajectory can be seen. However, this is unproblematic since measured motor angles are used together with the current torques for the determination of stiffness. This is done after the experiments. Thereby, a linear regression of torque~$\tau$ and angle~$\tilde{q}_\ind{m}$ is performed twice at each data point: once at the zero crossing from $\tilde{q}_\ind{m}{<}\SI{0}{\degree}$ to $\tilde{q}_\ind{m}{>}\SI{0}{\degree}$, illustrated in Fig.~\ref{fig:trajectory}(b), and a second time at the zero crossing in the opposite direction which is shown in Fig.~\ref{fig:trajectory}(c). The mean of both slopes is taken as measurement for the stiffness~$s_i$.

\begin{figure}[tbp]
	\vspace{1.5mm}
	\centerline{\includegraphics[width=\linewidth]{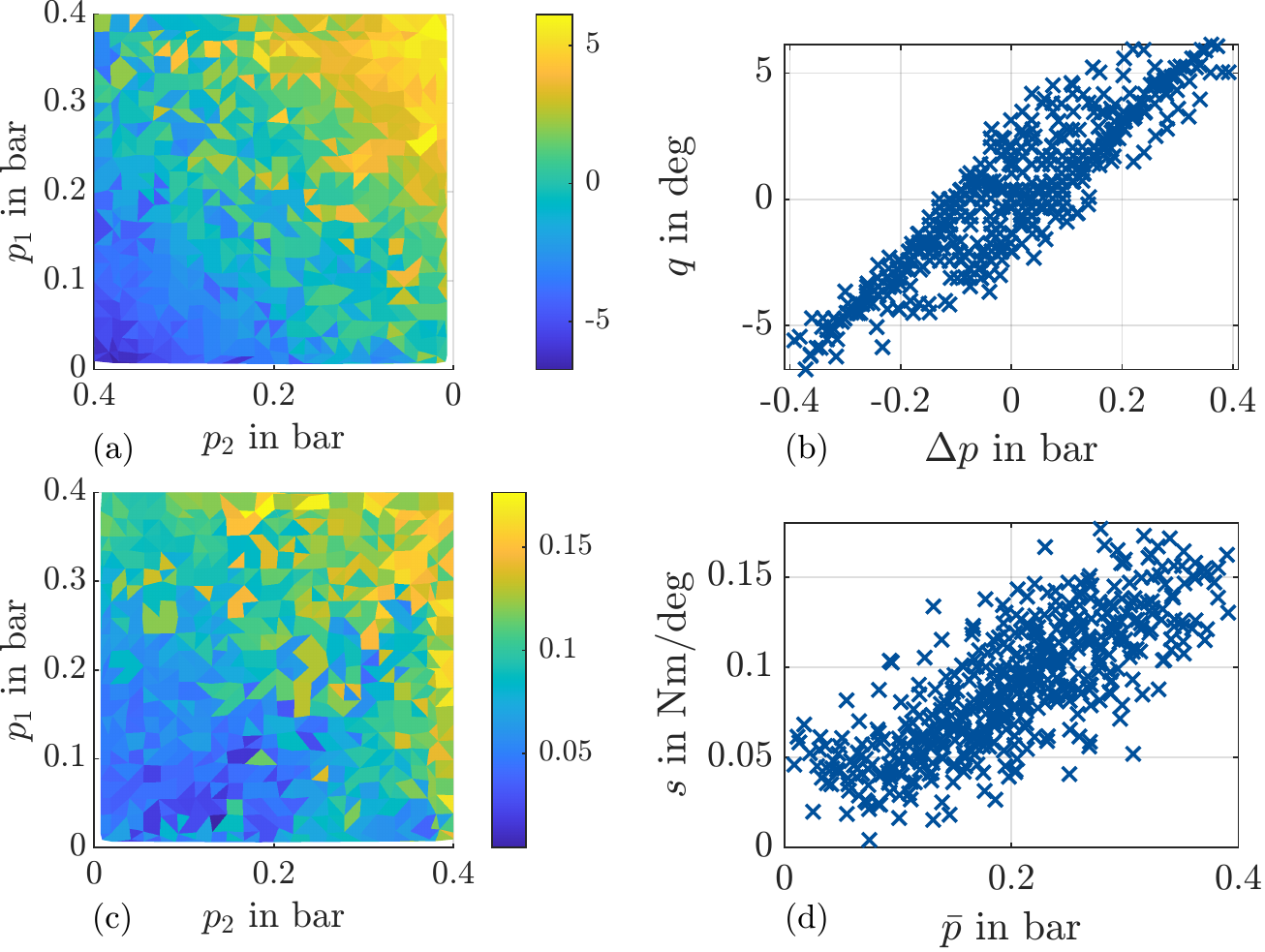}}
	\caption{Training data consisting of $n{=}518$ data points: (a)--(b) Joint angle plotted against the two bellows pressures and the pressure difference, respectively. (c)--(d) Joint stiffness plotted against the two bellows pressures and the mean pressure, respectively.}
	\label{fig:training_data}
	\vspace{-1.5mm}
\end{figure}

The total dataset consists of $n{=}518$ data points with pressures in the range of $\SI{0}{bar}{\leq} p_\diamond{\leq}\SI{0.4}{bar}$ relative to ambient pressure and is illustrated in Fig.~\ref{fig:training_data}. Pressure combinations are evenly distributed according to a full-factorial plan. Data recording took about five hours, whereby five bellows needed to be exchanged due to material failure. Comparable room temperatures were present during all experiments.
\subsection{Cross-Validation}\label{cv_sec}
\begin{figure}[tbp]
	\vspace{1.5mm}
	\centerline{\includegraphics[width=.9\linewidth]{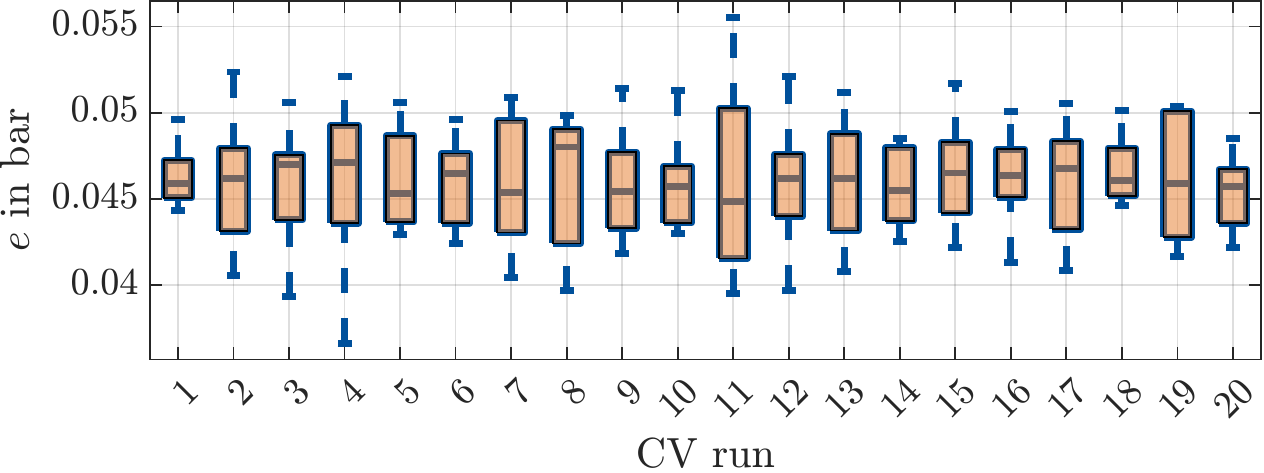}}
	\caption{CV results: Fold partitioning for each run is randomized and the MAE is calculated for $n_\ind{f}$ combinations of training and test data.}
	\label{fig:cv_results}
	\vspace{-1.5mm}
\end{figure}

To evaluate the performance of the trained predictor, $(n_\ind{f}{=}10)$-fold cross-validation (CV) is performed. To obtain representative results, this is repeated $n_\ind{CV}{=}20$ times, each run randomizing the division of the data into each fold. In all $n_\ind{CV}$ runs, the mean absolute error (MAE)
\e{e=\frac{1}{2n_\ind{test}}\sum_{j=1}^{n_\ind{test}} |\hat{y}_{\ind{I},j}-y_{\ind{I},j}|+|\hat{y}_{\ind{II},j}-y_{\ind{II},j}|}{mae}
is calculated for all $n_\ind{f}$ combinations of training and test dataset. Here $\hat{\mm{y}}_\ind{I},\,\hat{\mm{y}}_\ind{II}$ are the predicted pressures and $\mm{y}_\ind{I},\,\mm{y}_\ind{II}$ are the ground-truth of the pressures in the two bellows given desired joint angle and stiffness for $n_\ind{test}$ data points. Results of the cross-validation are shown in Fig.~\ref{fig:cv_results}. The mean absolute error is on average \SI{0.046}{bar} which corresponds to $11.5\%$ of the total pressure range ($0{-}\SI{0.4}{bar}$). This inaccuracy must be compensated by feedback control.
\subsection{Simultaneous Position and Stiffness Tracking}\label{control_sec}
The proposed learning-based position and stiffness control from Fig.~\ref{fig:blockschaltbild} was validated\footnote{Illustration of the procedure: \url{https://youtu.be/T9HzfOonz8U}} on the test bench from Fig.~\ref{fig:test_bench}. Seven different target joint angles $q_\ind{d}$ were commanded with a low joint stiffness and a high joint stiffness specified successively in each case. As mentioned above, the feedback loop affects the joint stiffness. High gains lead to a good position tracking but at the same time to an increase of the stiffness which is not desired. On the other hand, very low gains result in a low error compensation of the learned GPs and thus worse position tracking but also a low influence on the stiffness. The gains $K_\ind{P}{=}0.025\frac{\mathrm{bar}}{\mathrm{deg}}$ and $K_\ind{I}{=}0.05\frac{\mathrm{bar}}{\mathrm{s}\cdot\mathrm{deg}}$ were iteratively tuned to resolve this trade-off.

\begin{figure}[tbp]
	\vspace{1.5mm}
	\centerline{\includegraphics[width=\linewidth]{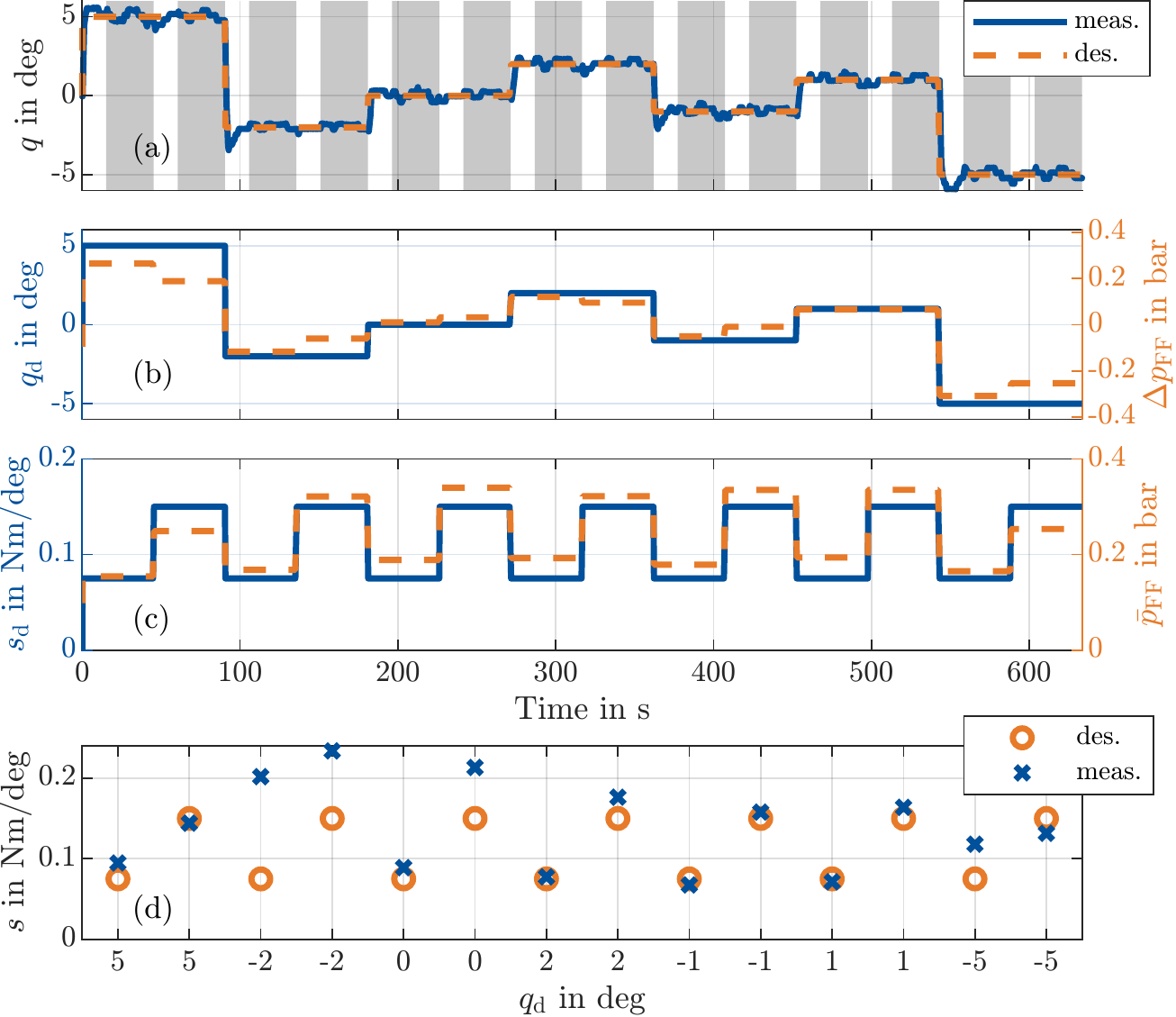}}
	\caption{Results of simultaneous position and stiffness control: (a) Desired and measured joint angle. (b) Desired joint angle and feedforward pressure difference $\Delta p_\ind{FF}{=}p_\ind{1,FF}{-}p_\ind{2,FF}$ resulting from GPs. (c) Desired joint stiffness and mean of the feedforward pressures $\bar{p}_\ind{FF}{=}(p_\ind{1,FF}{+}p_\ind{2,FF})/2$ resulting from GPs. (d) Comparison between desired and measured joint stiffness. Seven different desired joint angles with two different target stiffnesses were successively commanded. Gray areas indicate stiffness measurement with external disturbance (motor).}
	\label{fig:control_results}
	\vspace{-1.5mm}
\end{figure}

Position tracking results are shown in Fig.~\ref{fig:control_results}(a). In each target configuration $(q_\ind{d},s_\ind{d})$, the stiffness is measured using the approach from Fig.~\ref{fig:trajectory} indicated by the gray areas. Since the motor changes the joint angle of the actuator, deviation from the desired target angle occurs in these areas. To evaluate the position tracking, the areas without stiffness measurement should be considered. High accuracy (MAE of $\SI{0.34}{\degree}$ at encoder resolution of $\SI{0.35}{\degree}$) can be seen for all target angles. In Fig.~\ref{fig:control_results}(b)--(c), the predictions of the GPs are visualized. Accordingly, the pressure difference is adjusted to change the joint angle, whereas the average pressure is increased to achieve higher stiffness. The tracking of the stiffness is illustrated in Fig.~\ref{fig:control_results}(d). For each target angle, a low stiffness was first specified and then doubled. Solely for $q_\ind{d}{=}\SI{-2}{\degree}$ the measured joint stiffnesses $s$ are significantly larger than the commanded stiffnesses $s_\ind{d}$. This can be explained by the stiffening of the system due to the feedback loop. The relatively high loading $p_1$ applied to the first bellows during the previous configuration ($q_\ind{d}{=}\SI{5}{\degree}$) may also be a reason why the viscoelastic material has not recovered to its initial state. However, in all cases, the desired increase in joint stiffness is realized by the learning-based approach. Moreover, in ten out of 14 configurations, the desired stiffness is set with high accuracy.
\section{Conclusions}\label{conclusions}
This work presents a universal approach to learning-based position and stiffness control of VSAs. For this purpose, the fabrication and assembly of a modular soft pneumatic actuator is first presented. Using Gaussian processes, a model of the actuator is trained which has an error of $11.5\%$ of the total pressure range and is used for feedforward control. The presented data-driven control allows continuous specification of joint angle and joint stiffness and is applicable to \textit{arbitrary actuator designs} with variable joint stiffness. Future work arises from this work. Soft actuators suffer from material fatigue and thus show a change in mechanical behavior with increasing number of cycles \cite{Klute.1998}. The use of an online learning approach would counter this by continuously updating the learned actuator model. However, this requires an online measurement of the joint stiffness which has to be done during operation without motor and torque sensor. For this purpose, recent studies in the field of stiffness observers \cite{Fagiolini.2020,Trumic.2022} are a promising approach. Feedback of the joint stiffness would also be possible with such an observer. Further, extending the method to multi-DoF systems is required for data-driven control of soft robots with variable stiffness.



\section*{Acknowledgment}
The authors acknowledge the support of this project by the German Research Foundation (Deutsche Forschungsgemeinschaft) under grant number 433586601.
\bibliographystyle{IEEEtran}
\bibliography{literatur}

\end{document}

%% file: images/cover.pdf_tex
\begingroup%
  \makeatletter%
  \providecommand\color[2][]{%
    \errmessage{(Inkscape) Color is used for the text in Inkscape, but the package 'color.sty' is not loaded}%
    \renewcommand\color[2][]{}%
  }%
  \providecommand\transparent[1]{%
    \errmessage{(Inkscape) Transparency is used (non-zero) for the text in Inkscape, but the package 'transparent.sty' is not loaded}%
    \renewcommand\transparent[1]{}%
  }%
  \providecommand\rotatebox[2]{#2}%
  \newcommand*\fsize{\dimexpr\f@size pt\relax}%
  \newcommand*\lineheight[1]{\fontsize{\fsize}{#1\fsize}\selectfont}%
  \ifx\svgwidth\undefined%
    \setlength{\unitlength}{202.97839253bp}%
    \ifx\svgscale\undefined%
      \relax%
    \else%
      \setlength{\unitlength}{\unitlength * \real{\svgscale}}%
    \fi%
  \else%
    \setlength{\unitlength}{\svgwidth}%
  \fi%
  \global\let\svgwidth\undefined%
  \global\let\svgscale\undefined%
  \makeatother%
  \begin{picture}(1,0.58358339)%
    \lineheight{1}%
    \setlength\tabcolsep{0pt}%
    \put(0,0){\includegraphics[width=\unitlength,page=1]{cover.pdf}}%
    \put(-0.00112196,0.55910487){\makebox(0,0)[lt]{\lineheight{1.25}\smash{\begin{tabular}[t]{l}\scriptsize{(a)}\end{tabular}}}}%
    \put(0.51186544,0.55910487){\makebox(0,0)[lt]{\lineheight{1.25}\smash{\begin{tabular}[t]{l}\scriptsize{(b)}\end{tabular}}}}%
    \put(0,0){\includegraphics[width=\unitlength,page=2]{cover.pdf}}%
    \put(0.00677868,0.45025242){\makebox(0,0)[lt]{\lineheight{1.25}\smash{\begin{tabular}[t]{l}\scriptsize{variable}\end{tabular}}}}%
    \put(0.03041367,0.41110415){\makebox(0,0)[lt]{\lineheight{1.25}\smash{\begin{tabular}[t]{l}\scriptsize{angle}\end{tabular}}}}%
    \put(-0.00016304,0.15370622){\makebox(0,0)[lt]{\lineheight{1.25}\smash{\begin{tabular}[t]{l}\scriptsize{variable}\end{tabular}}}}%
    \put(-0.00016304,0.11455871){\makebox(0,0)[lt]{\lineheight{1.25}\smash{\begin{tabular}[t]{l}\scriptsize{stiffness}\end{tabular}}}}%
    \put(0,0){\includegraphics[width=\unitlength,page=3]{cover.pdf}}%
  \end{picture}%
\endgroup%

%% file: images/construction_view.pdf_tex
\begingroup%
  \makeatletter%
  \providecommand\color[2][]{%
    \errmessage{(Inkscape) Color is used for the text in Inkscape, but the package 'color.sty' is not loaded}%
    \renewcommand\color[2][]{}%
  }%
  \providecommand\transparent[1]{%
    \errmessage{(Inkscape) Transparency is used (non-zero) for the text in Inkscape, but the package 'transparent.sty' is not loaded}%
    \renewcommand\transparent[1]{}%
  }%
  \providecommand\rotatebox[2]{#2}%
  \newcommand*\fsize{\dimexpr\f@size pt\relax}%
  \newcommand*\lineheight[1]{\fontsize{\fsize}{#1\fsize}\selectfont}%
  \ifx\svgwidth\undefined%
    \setlength{\unitlength}{222.42623629bp}%
    \ifx\svgscale\undefined%
      \relax%
    \else%
      \setlength{\unitlength}{\unitlength * \real{\svgscale}}%
    \fi%
  \else%
    \setlength{\unitlength}{\svgwidth}%
  \fi%
  \global\let\svgwidth\undefined%
  \global\let\svgscale\undefined%
  \makeatother%
  \begin{picture}(1,0.68482263)%
    \lineheight{1}%
    \setlength\tabcolsep{0pt}%
    \put(0,0){\includegraphics[width=\unitlength,page=1]{construction_view.pdf}}%
    \put(0.15710594,0.55805487){\makebox(0,0)[lt]{\lineheight{1.25}\smash{\begin{tabular}[t]{l}\scriptsize platform\end{tabular}}}}%
    \put(0.17193046,0.52409482){\makebox(0,0)[lt]{\lineheight{1.25}\smash{\begin{tabular}[t]{l}\scriptsize (rigid)\end{tabular}}}}%
    \put(0.0687593,0.48552863){\makebox(0,0)[lt]{\lineheight{1.25}\smash{\begin{tabular}[t]{l}\scriptsize sealing\end{tabular}}}}%
    \put(0.07683989,0.45152821){\makebox(0,0)[lt]{\lineheight{1.25}\smash{\begin{tabular}[t]{l}\scriptsize (soft)\end{tabular}}}}%
    \put(0.02829695,0.2706133){\makebox(0,0)[lt]{\lineheight{1.25}\smash{\begin{tabular}[t]{l}\scriptsize (soft)\end{tabular}}}}%
    \put(-0.00001519,0.30506248){\makebox(0,0)[lt]{\lineheight{1.25}\smash{\begin{tabular}[t]{l}\scriptsize membrane\end{tabular}}}}%
    \put(0,0){\includegraphics[width=\unitlength,page=2]{construction_view.pdf}}%
    \put(0.81202842,0.43771639){\makebox(0,0)[lt]{\lineheight{1.25}\smash{\begin{tabular}[t]{l}\scriptsize bellows\end{tabular}}}}%
    \put(0.84574749,0.38262871){\makebox(0,0)[lt]{\lineheight{1.25}\smash{\begin{tabular}[t]{l}\scriptsize rigid joint\end{tabular}}}}%
    \put(0.86664845,0.29252143){\makebox(0,0)[lt]{\lineheight{1.25}\smash{\begin{tabular}[t]{l}\scriptsize magnet\end{tabular}}}}%
    \put(0,0){\includegraphics[width=\unitlength,page=3]{construction_view.pdf}}%
    \put(0.8936237,0.24889738){\makebox(0,0)[lt]{\lineheight{1.25}\smash{\begin{tabular}[t]{l}\scriptsize Hall\end{tabular}}}}%
    \put(0.46201976,0.65352592){\makebox(0,0)[lt]{\lineheight{1.25}\smash{\begin{tabular}[t]{l}\scriptsize (b)\end{tabular}}}}%
    \put(0.82900532,0.51091596){\makebox(0,0)[lt]{\lineheight{1.25}\smash{\begin{tabular}[t]{l}\scriptsize tube\end{tabular}}}}%
    \put(0,0){\includegraphics[width=\unitlength,page=4]{construction_view.pdf}}%
    \put(0.88013607,0.22192213){\makebox(0,0)[lt]{\lineheight{1.25}\smash{\begin{tabular}[t]{l}\scriptsize sensor\end{tabular}}}}%
    \put(0.78179863,0.48394071){\makebox(0,0)[lt]{\lineheight{1.25}\smash{\begin{tabular}[t]{l}\scriptsize connections\end{tabular}}}}%
    \put(0,0){\includegraphics[width=\unitlength,page=5]{construction_view.pdf}}%
    \put(0.82618561,0.62655068){\makebox(0,0)[lt]{\lineheight{1.25}\smash{\begin{tabular}[t]{l}\scriptsize upper\end{tabular}}}}%
    \put(0.82618561,0.59283161){\makebox(0,0)[lt]{\lineheight{1.25}\smash{\begin{tabular}[t]{l}\scriptsize frame\end{tabular}}}}%
    \put(0,0){\includegraphics[width=\unitlength,page=6]{construction_view.pdf}}%
    \put(0.79246654,0.05332692){\makebox(0,0)[lt]{\lineheight{1.25}\smash{\begin{tabular}[t]{l}\scriptsize lower\end{tabular}}}}%
    \put(0.79246654,0.02635166){\makebox(0,0)[lt]{\lineheight{1.25}\smash{\begin{tabular}[t]{l}\scriptsize frame\end{tabular}}}}%
    \put(0.03716014,0.65352592){\makebox(0,0)[lt]{\lineheight{1.25}\smash{\begin{tabular}[t]{l}\scriptsize (a)\end{tabular}}}}%
    \put(0.85990463,0.34520738){\makebox(0,0)[lt]{\lineheight{1.25}\smash{\begin{tabular}[t]{l}\scriptsize  (1-DoF)\end{tabular}}}}%
  \end{picture}%
\endgroup%

%% file: images/blockschaltbild.pdf_tex
\begingroup%
  \makeatletter%
  \providecommand\color[2][]{%
    \errmessage{(Inkscape) Color is used for the text in Inkscape, but the package 'color.sty' is not loaded}%
    \renewcommand\color[2][]{}%
  }%
  \providecommand\transparent[1]{%
    \errmessage{(Inkscape) Transparency is used (non-zero) for the text in Inkscape, but the package 'transparent.sty' is not loaded}%
    \renewcommand\transparent[1]{}%
  }%
  \providecommand\rotatebox[2]{#2}%
  \newcommand*\fsize{\dimexpr\f@size pt\relax}%
  \newcommand*\lineheight[1]{\fontsize{\fsize}{#1\fsize}\selectfont}%
  \ifx\svgwidth\undefined%
    \setlength{\unitlength}{219.13875891bp}%
    \ifx\svgscale\undefined%
      \relax%
    \else%
      \setlength{\unitlength}{\unitlength * \real{\svgscale}}%
    \fi%
  \else%
    \setlength{\unitlength}{\svgwidth}%
  \fi%
  \global\let\svgwidth\undefined%
  \global\let\svgscale\undefined%
  \makeatother%
  \begin{picture}(1,0.34934077)%
    \lineheight{1}%
    \setlength\tabcolsep{0pt}%
    \put(0,0){\includegraphics[width=\unitlength,page=1]{blockschaltbild.pdf}}%
    \put(0.18852173,0.04724504){\makebox(0,0)[lt]{\lineheight{1.25}\smash{\begin{tabular}[t]{l}\scriptsize PI\end{tabular}}}}%
    \put(0,0){\includegraphics[width=\unitlength,page=2]{blockschaltbild.pdf}}%
    \put(0.11219241,0.29914623){\makebox(0,0)[lt]{\lineheight{1.25}\smash{\begin{tabular}[t]{l}\scriptsize GP prediction\end{tabular}}}}%
    \put(0,0){\includegraphics[width=\unitlength,page=3]{blockschaltbild.pdf}}%
    \put(0.01146173,0.29924179){\makebox(0,0)[lt]{\lineheight{1.25}\smash{\begin{tabular}[t]{l}\scriptsize$q_\mathrm{d}$\end{tabular}}}}%
    \put(0.01146173,0.34567468){\makebox(0,0)[lt]{\lineheight{1.25}\smash{\begin{tabular}[t]{l}\scriptsize$s_\mathrm{d}$\end{tabular}}}}%
    \put(0,0){\includegraphics[width=\unitlength,page=4]{blockschaltbild.pdf}}%
    \put(0.92762155,0.1958898){\makebox(0,0)[lt]{\lineheight{1.25}\smash{\begin{tabular}[t]{l}\scriptsize$q$\end{tabular}}}}%
    \put(0.2109021,0.24576758){\makebox(0,0)[lt]{\lineheight{1.25}\smash{\begin{tabular}[t]{l}\scriptsize$\mm{p}_\mathrm{FF}$\end{tabular}}}}%
    \put(0.21090091,0.11216619){\makebox(0,0)[lt]{\lineheight{1.25}\smash{\begin{tabular}[t]{l}\scriptsize$\Delta p_\mathrm{FB}$\end{tabular}}}}%
    \put(0.32066437,0.06710688){\makebox(0,0)[lt]{\lineheight{1.25}\smash{\begin{tabular}[t]{l}\scriptsize$-$\end{tabular}}}}%
    \put(0.27697542,0.198851){\makebox(0,0)[lt]{\lineheight{1.25}\smash{\begin{tabular}[t]{l}\scriptsize$\mm{p}_\mathrm{d}$\end{tabular}}}}%
    \put(0,0){\includegraphics[width=\unitlength,page=5]{blockschaltbild.pdf}}%
    \put(0.17117625,0.17230032){\makebox(0,0)[lt]{\lineheight{1.25}\smash{\begin{tabular}[t]{l}\scriptsize$\mm{g}(\cdot)$\end{tabular}}}}%
    \put(0,0){\includegraphics[width=\unitlength,page=6]{blockschaltbild.pdf}}%
    \put(0.48620089,0.198851){\makebox(0,0)[lt]{\lineheight{1.25}\smash{\begin{tabular}[t]{l}\scriptsize$\mm{p}$\end{tabular}}}}%
    \put(0,0){\includegraphics[width=\unitlength,page=7]{blockschaltbild.pdf}}%
    \put(0.36671193,0.17230032){\makebox(0,0)[lt]{\lineheight{1.25}\smash{\begin{tabular}[t]{l}\scriptsize valves\end{tabular}}}}%
    \put(0,0){\includegraphics[width=\unitlength,page=8]{blockschaltbild.pdf}}%
    \put(0.75200167,0.25087507){\makebox(0,0)[lt]{\lineheight{1.25}\smash{\begin{tabular}[t]{l}\scriptsize$q$\end{tabular}}}}%
    \put(0,0){\includegraphics[width=\unitlength,page=9]{blockschaltbild.pdf}}%
    \put(0.66301786,0.25087507){\makebox(0,0)[lt]{\lineheight{1.25}\smash{\begin{tabular}[t]{l}\scriptsize$p_1$\end{tabular}}}}%
    \put(0.84783212,0.25087507){\makebox(0,0)[lt]{\lineheight{1.25}\smash{\begin{tabular}[t]{l}\scriptsize$p_2$\end{tabular}}}}%
    \put(0,0){\includegraphics[width=\unitlength,page=10]{blockschaltbild.pdf}}%
  \end{picture}%
\endgroup%

%% file: images/test_bench.pdf_tex
\begingroup%
  \makeatletter%
  \providecommand\color[2][]{%
    \errmessage{(Inkscape) Color is used for the text in Inkscape, but the package 'color.sty' is not loaded}%
    \renewcommand\color[2][]{}%
  }%
  \providecommand\transparent[1]{%
    \errmessage{(Inkscape) Transparency is used (non-zero) for the text in Inkscape, but the package 'transparent.sty' is not loaded}%
    \renewcommand\transparent[1]{}%
  }%
  \providecommand\rotatebox[2]{#2}%
  \newcommand*\fsize{\dimexpr\f@size pt\relax}%
  \newcommand*\lineheight[1]{\fontsize{\fsize}{#1\fsize}\selectfont}%
  \ifx\svgwidth\undefined%
    \setlength{\unitlength}{249.00652865bp}%
    \ifx\svgscale\undefined%
      \relax%
    \else%
      \setlength{\unitlength}{\unitlength * \real{\svgscale}}%
    \fi%
  \else%
    \setlength{\unitlength}{\svgwidth}%
  \fi%
  \global\let\svgwidth\undefined%
  \global\let\svgscale\undefined%
  \makeatother%
  \begin{picture}(1,0.73156033)%
    \lineheight{1}%
    \setlength\tabcolsep{0pt}%
    \put(0,0){\includegraphics[width=\unitlength,page=1]{test_bench.pdf}}%
    \put(0.14414033,0.09224461){\makebox(0,0)[lt]{\lineheight{1.25}\smash{\begin{tabular}[t]{l}\scriptsize Ethernet\end{tabular}}}}%
    \put(0.16730157,0.38109355){\makebox(0,0)[lt]{\lineheight{1.25}\smash{\begin{tabular}[t]{l}\scriptsize$q$\end{tabular}}}}%
    \put(0,0){\includegraphics[width=\unitlength,page=2]{test_bench.pdf}}%
    \put(0.56349351,0.53467281){\makebox(0,0)[lt]{\lineheight{1.25}\smash{\begin{tabular}[t]{l}\scriptsize VSA\end{tabular}}}}%
    \put(0.54742566,0.31895322){\makebox(0,0)[lt]{\lineheight{1.25}\smash{\begin{tabular}[t]{l}\scriptsize{valves}\end{tabular}}}}%
    \put(0,0){\includegraphics[width=\unitlength,page=3]{test_bench.pdf}}%
    \put(0.49456472,0.09172226){\makebox(0,0)[lt]{\lineheight{1.25}\smash{\begin{tabular}[t]{l}\scriptsize supply unit\end{tabular}}}}%
    \put(0,0){\includegraphics[width=\unitlength,page=4]{test_bench.pdf}}%
    \put(0.22748813,0.31017028){\makebox(0,0)[lt]{\lineheight{1.25}\smash{\begin{tabular}[t]{l}\scriptsize$\mm{p}$\end{tabular}}}}%
    \put(0.33881509,0.28456091){\color[rgb]{0,0.31372549,0.60784314}\makebox(0,0)[lt]{\lineheight{1.25}\smash{\begin{tabular}[t]{l}\scriptsize$2$\end{tabular}}}}%
    \put(0.35230259,0.18594622){\color[rgb]{0,0.31372549,0.60784314}\makebox(0,0)[lt]{\lineheight{1.25}\smash{\begin{tabular}[t]{l}\scriptsize$3$\end{tabular}}}}%
    \put(0.29600802,0.18594622){\color[rgb]{0,0.31372549,0.60784314}\makebox(0,0)[lt]{\lineheight{1.25}\smash{\begin{tabular}[t]{l}\scriptsize$1$\end{tabular}}}}%
    \put(0.07619453,0.02481042){\makebox(0,0)[lt]{\lineheight{1.25}\smash{\begin{tabular}[t]{l}\scriptsize DEV-PC\end{tabular}}}}%
    \put(0.08857042,0.15549967){\makebox(0,0)[lt]{\lineheight{1.25}\smash{\begin{tabular}[t]{l}\scriptsize RT-PC\end{tabular}}}}%
    \put(0.06922511,0.28546211){\makebox(0,0)[lt]{\lineheight{1.25}\smash{\begin{tabular}[t]{l}\scriptsize EtherCAT\end{tabular}}}}%
    \put(0.14414033,0.22079154){\makebox(0,0)[lt]{\lineheight{1.25}\smash{\begin{tabular}[t]{l}\scriptsize EtherLab\end{tabular}}}}%
    \put(0.03420612,0.62961829){\makebox(0,0)[lt]{\lineheight{1.25}\smash{\begin{tabular}[t]{l}\scriptsize ROS shared memory\end{tabular}}}}%
    \put(0.41215176,0.26071587){\color[rgb]{0,0.31372549,0.60784314}\makebox(0,0)[lt]{\lineheight{1.25}\smash{\begin{tabular}[t]{l}\tiny P\end{tabular}}}}%
    \put(0.55278088,0.1442231){\makebox(0,0)[lt]{\lineheight{1.25}\smash{\begin{tabular}[t]{l}\scriptsize$\mm{p}_\mathrm{d}$\end{tabular}}}}%
    \put(0,0){\includegraphics[width=\unitlength,page=5]{test_bench.pdf}}%
    \put(0.50296742,0.28456089){\color[rgb]{0,0.31372549,0.60784314}\makebox(0,0)[lt]{\lineheight{1.25}\smash{\begin{tabular}[t]{l}\scriptsize$2$\end{tabular}}}}%
    \put(0.51645484,0.18594621){\color[rgb]{0,0.31372549,0.60784314}\makebox(0,0)[lt]{\lineheight{1.25}\smash{\begin{tabular}[t]{l}\scriptsize$3$\end{tabular}}}}%
    \put(0.46016027,0.18594621){\color[rgb]{0,0.31372549,0.60784314}\makebox(0,0)[lt]{\lineheight{1.25}\smash{\begin{tabular}[t]{l}\scriptsize$1$\end{tabular}}}}%
    \put(0.57630409,0.26071585){\color[rgb]{0,0.31372549,0.60784314}\makebox(0,0)[lt]{\lineheight{1.25}\smash{\begin{tabular}[t]{l}\tiny P\end{tabular}}}}%
    \put(0,0){\includegraphics[width=\unitlength,page=6]{test_bench.pdf}}%
    \put(0.38202239,0.6062072){\makebox(0,0)[lt]{\lineheight{1.25}\smash{\begin{tabular}[t]{l}\scriptsize torque sensor\end{tabular}}}}%
    \put(0,0){\includegraphics[width=\unitlength,page=7]{test_bench.pdf}}%
    \put(0.41816588,0.68828337){\makebox(0,0)[lt]{\lineheight{1.25}\smash{\begin{tabular}[t]{l}\scriptsize motor\end{tabular}}}}%
    \put(0.01966221,0.13547608){\makebox(0,0)[lt]{\lineheight{1.25}\smash{\begin{tabular}[t]{l}\scriptsize$\tau$\end{tabular}}}}%
    \put(0.07034981,0.58242897){\makebox(0,0)[lt]{\lineheight{1.25}\smash{\begin{tabular}[t]{l}\scriptsize USB/Ethernet\end{tabular}}}}%
    \put(0,0){\includegraphics[width=\unitlength,page=8]{test_bench.pdf}}%
    \put(0.19742147,0.71241024){\makebox(0,0)[lt]{\lineheight{1.25}\smash{\begin{tabular}[t]{l}\scriptsize$q_\mathrm{m,d}$\end{tabular}}}}%
    \put(0,0){\includegraphics[width=\unitlength,page=9]{test_bench.pdf}}%
    \put(0.05284655,0.38109355){\makebox(0,0)[lt]{\lineheight{1.25}\smash{\begin{tabular}[t]{l}\scriptsize$q_\mathrm{m}$\end{tabular}}}}%
    \put(0,0){\includegraphics[width=\unitlength,page=10]{test_bench.pdf}}%
  \end{picture}%
\endgroup%